\documentclass[conference]{IEEEtran}
\usepackage{pifont}
\usepackage{amsfonts}
\usepackage{amssymb}
\usepackage{amsmath}
\usepackage{epsfig}
\usepackage{color}
\usepackage{cite}

\newtheorem{theorem}{Theorem}
\newtheorem{corollary}{Corollary}
\newtheorem{proposition}{Proposition}

\newtheorem{definition}{Definition}

\newtheorem{lemma}{Lemma}

\DeclareMathOperator{\Tr}{Tr}

\def\QED{\mbox{\rule[0pt]{1.5ex}{1.5ex}}}

\begin{document}

\title{On-the-fly Uplink Training and Pilot Code Sequence Design for Cellular Networks}
\author{\IEEEauthorblockN{Zekun Zhang$^*$, Chenwei Wang$^{\dagger}$, Haralabos Papadopoulos$^{\dagger}$}\\
\IEEEauthorblockA{$^*$Dept. of Electricla and Computer Engineering, Utah State University, Logan, UT 84322\\
$^{\dagger}$DOCOMO Innovations, Inc., Palo Alto, CA 94304\\
$^*$zekun.zhang.z@ieee.org, $^{\dagger}$\{cwang, hpapadopoulos\}@docomoinnovations.com }}


\maketitle

\begin{abstract}
Cellular networks of massive MIMO base-stations employing TDD/OFDM and relying on uplink training for both downlink and uplink transmission are viewed as an attractive candidate for 5G deployments, as they promise high area spectral and energy efficiencies with relatively simple low-latency operation. We investigate the use of non-orthogonal uplink pilot designs as a means for improving the area spectral efficiency in the downlink of such massive MIMO cellular networks. We develop a class of pilot designs that are locally orthogonal within each cell, while maintaining low inner-product properties between codes in different cells. Using channel estimates provided by observations on these codes, each cell independently serves its locally active users with MU-MIMO transmission that is also designed to mitigate interference to  a subset of ``strongly interfered'' out-of-cell users.  As our simulation-based analysis shows, such cellular operation based on the proposed codes yields user-rate CDF improvement with respect to conventional operation, which can be exploited to improve cell and/or cell-throughput performance.

\end{abstract}

\allowdisplaybreaks

\section{Introduction}\label{sec:intro}

Since its introduction in \cite{Marzetta-2010},
massive MIMO  has been increasingly viewed as a key technological and economic driver for 5G and beyond deployments. Cellular massive MIMO operated in TDD/OFDM (Time Division Duplexing/Orthogonal Frequency Division Multiplexing) and relying on training in the uplink for both downlink and uplink transmission offers unique operational advantages.  Aside form maintaining the simplicity of cellular operation, uplink training readily provides user-channel estimates to the infrastructure readily and with low overheads, the channel between a user antenna and all the infrastructure antennas can be obtained by a single uplink pilot transmission. This is in contrast to  conventional FDD cellular operation, whereby downlink reference signals are used to obtain channel estimates at the user equipments (UEs), and these estimates are then fed back through the uplink to the infrastructure. A brute-force approach to learning the channel between all base-station (BS) antennas and a UE scales linearly with the number of antennas. Recently, a lot of effort has been placed to improve the efficiency of downlink training schemes and new more resource-efficient schemes have been designed that exploit the physics of the propagation channels, e.g., \cite{Ansuman-JSDM}.

Many new 5G deployments are expected at higher-frequency bands included mmWave, where abundant bandwidth is available. Due the radio propagation characteristics at these higher frequencies, coverage from a single BS is expected to be shorter, substantially spottier and intermittent than at lower frequency bands \cite{ICC2018}. As a result, traditional cell planning with frequency reuse no longer applies, and denser inherently irregular deployments are required to provide sufficient coverage in these bands.

To cope with such dense irregular deployments, a new class of non-orthogonal pilot codes were proposed in \cite{ozgun-icc-2016} for networks of massive MIMO radio remote head (RRH) networks, where users across the network share pilot dimensions in such a way that they can be opportunistically served by a subset of nearby RRHs. Referred to as Spotlight, it exploits ``overloaded'' pilot reuse (many users are aligned on a UL pilot dimension), and exploits  fast user detection at each RRH based on a simple binary energy detection scheme.

In this paper, we focus on a TDD/OFDM system that exploits reciprocity-based training, as in e.g., \cite{Marzetta-2010, ozgun-icc-2016}. As in \cite{Marzetta-2010}, but in contrast to \cite{ozgun-icc-2016}, we focus on cellular operation. In particular, we assume slotted downlink transmission over  resource blocks (RBs) on the OFDM plane, and consider a quasi-static channel model according to which the user channels remain constant within all resource elements (REs) comprising a single RB (or slot). In each slot, we assume that each cell schedules  $K$ users.  As in  \cite{Marzetta-2010}, all scheduled users transmit uplink pilots over a set of $Q$ REs allocated for uplink training. Based on observations collected over these $Q$ REs, BSs learn their user-channels and subsequently serve them in the remaining REs in the slot.  We assume that each BS knows the uplink pilot codes used in all (or, in practice, all nearby) base stations but not the identities of scheduled users.

We focus on the use of appropriately designed non-orthogonal uplink pilot codes as a means for increasing the area spectral efficiency with respect to conventional operation  \cite{Marzetta-2010}.  The operation we propose exploits the findings in \cite{ICC2018}, according to which there exist pilot designs that allow each BS to estimate all the large-scale gains of all scheduled users (at all nearby BSs)  provided the number of scheduled users does not exceed $Q^2$. Indeed, this enables each BS to identify out-of-cell users that have sufficiently strong channels to the BS, and can tune its downlink transmission so as to mitigate interference to these users.

We develop a class of new non-orthogonal uplink pilot designs, according to which any pair of pilot codewords used within a cell are orthogonal,  while the magnitude of the inner product of any pair of pilot codewords used in different cells is small (the optimality condition of the new design is also investigated for a number of cases). Subsequently, assuming the use of these codes for uplink training, we consider the use of cellular MU-MIMO transmission which also  mitigates interference to the strongest out-of-cell users. Finally, we evaluate the viability of the proposed schemes by examining the achievable user-rates they provide over a sample 7-cell simulation scenario, and comparing them against their conventional system counterparts.

{\it Notations:} We use $a,~{\bf a},~{\bf A}$ to denote scalar, vector and matrix respectively. For matrix ${\bf A}$, we denote by ${\bf a}_k$ its $k$-th column; and for vector ${\bf a}$, we denote by ${\bf a}(k)$ its $k$-th entry. For matrix ${\bf A}$, its $k$-th row and $k$-th column are represented by ${\bf A}(k,:)$ and ${\bf A}(:,k)$, respectively, and its $m$-th row $n$-th column is represented by $({\bf A})_{m,n}$. Moreover, $|\cdot|$, $\|\cdot\|$, ``$\otimes$", and $\Tr(\cdot)$ denote the absolute-value operator, the norm-2 operator, the Kronecker-product operator and the Trace operator, respectively. Finally, ${\bf I}_Q$, ${\bf 1}_Q$ represent the $Q\times Q$ identity matrix, and the $Q\times Q$ all one matrix with all entries equal to one, respectively.

\section{System Model}

We consider a cellular network of $J$ cells, and the BS in each cell is equipped with $M$-antennas serving $K$ single-antenna UEs. We denote the BS in cell $j$ by $\text{BS}_j$, and the $k$-th user in cell $j$ by $\text{UE}_{jk}$, where $j = 1, \dotsb, J$ and $k = 1, \dotsb, K$. We assume the system is operated on TDD/OFDM. We employ a quasi-static channel model where each user channel remains constant within a RB comprising a set of REs within the user-channel coherence time and bandwidth. The channel between $\text{UE}_{jk}$ and $\text{BS}_i$ is denoted by an $M \times 1$ column vector  ${\bf h}_{ijk} = [h_{ijk}(1), h_{ijk}(2), \cdots, h_{ijk}(M)]^T \sim \mathcal{CN}({\bf 0}, g_{ijk}{\bf I}_M)$, where $g_{ijk}$ represents its large-scale channel-fading gain. In this work, we consider reciprocity-based training for downlink transmission over a generic RB, according to which the downlink user channels are learned at the BS via uplink user pilots training within that RB.

The entire communication consists of uplink transmission and downlink transmission phases. We assume both uplink and downlink transmissions are synchronized throughout the network. In the uplink phase, each active user sends pre-determined uplink pilot signals for the BS to estimate each user's channel. We assume that $Q$ REs are allocated for uplink pilot transmission in the RB, and each $\text{UE}_{jk}$ broadcasts a $Q\times 1$ pilot sequence, denoted by ${\bf p}_{jk}$. This pilot sequence has been pre-assigned to $\text{UE}_{jk}$, and the collection of $\{ {\bf p}_{jk} \}$'s are also known to all (sufficiently nearby) cells in the network. With $Q$ REs for training, the BS can learn the complete $M\times 1$ channel for at most $Q$ users. Therefore in this work we assume $K$ is the number of users that each BS serves in each RB\footnote{In practice, the actual number of users associated with one BS can definitely be greater than $Q$, where only certain portion of the users are scheduled for training and transmission within each coherence time and bandwidth following some scheduling scheme. Such problems are well studied under the topic of user scheduling and thus out of the scope of this work.} and $K \leq Q$. When all the users simultaneously train their channels, the received signal at BS$_i$ is given by:
\begin{equation}
{\bf Y}_i = \sum_{j=1}^J \sum_{k = 1}^K {\bf p}_{jk} {\bf h}_{ijk}^{T} + {\bf W}_i,\label{eqn:ul}
\end{equation}
where ${\bf W}_i$ is the noise matrix, and each entry is i.i.d. and follows $\mathcal{CN}(0,\sigma^2_u)$. For brevity, we assume the unit equipment transmit power for uplink training, i.e., $\|{\bf p}_{jk}\|^2 = 1$, and absorbing the actual transmission power into the noise power, i.e., $\sigma^2_u=\sigma^2_w/\rho_u$ where $\sigma^2_w$ is the thermal noise power and $\rho_u$ is the transmission power of each user. Note that both ${\bf Y}_i$ and ${\bf W}_i$ are $Q \times M$ matrices.

Prior to downlink transmission, BS$_i$ uses ${\bf Y}_i$ to first estimate the channels $\hat{{\bf h}}_{iik}$ of its cell's own users.
Subsequently, BS$_i$ computes a $M\times 1$ precoding vector ${\bf v}_{ik}$ to each intended user UE$_{ik}$. In this work we restrict our attention to widely used zero-forcing beamforming (ZFBF), so as to achieve MU-MIMO benefits. Denoting by $y_{ik}$ the received downlink signal at UE$_{ik}$ in cell $i$, we have:
\begin{equation}
y_{ik} = \sum_{j=1}^J \sum_{k=1}^K {\bf v}_{jk} {\bf h}_{ijk}^Ts_{jk} + z_{ik}\label{eqn:dl}
\end{equation}
where $s_{ijk}$, $\mathbb{E}[||s_{jk}||^2] = 1$, represents the independent data stream for UE$_{jk}$, and $z_{ik}$ is i.i.d. and follows $\mathcal{CN}(0, \sigma^2_d)$ where $\sigma^2_d = K\sigma^2_w/\rho_d$ represents the normalized Gaussian noise by assuming equal power allocation for each user where $\rho_d$ is the transmission power of each BS.

In this work, we investigate the design and use of non-orthogonal pilot code sequences for training the users' channels. In sections that follow, we propose a novel uplink training scheme and non-orthogonal pilots, and then assess the resulting downlink data-rate performance.

\section{The Proposed Scheme}

In this section, we introduce the complete process of using non-orthogonal pilot codes for channel estimation. The proposed scheme consists of two steps. First, the BS estimates the large-scale channel gains of nearby users using the scheme recently introduced by Wang {\em et al.} in \cite{ICC2018}. Subsequently, using these estimated large-scale channel gains, the BS estimates the small-scale fading of the subset of the user-channels for which their large-scale channel gains significantly larger than zero, i.e., the users whose channels to the given BS are sufficiently strong. Finally, each BS exploits the estimated channels for beamforming. In particular, it tries to mitigate interference to nearby out-of-cell users, while also leveraging MU-MIMO to serve multiple users in the downlink.

\subsection{Uplink: Large-Scale Channel Estimation}

Rewriting (\ref{eqn:ul}) in a compact matrix form, we have:
\small
\begin{eqnarray}
{\bf Y}_i &\!\!\!\!=\!\!\!\!& {\bf P} {\bf H}_i + {\bf W}_i,\label{received_signal_original}\\
\textrm{where}~~~~{\bf P} &\!\!\!\!=\!\!\!\!& [{\bf p}_{11},~{\bf p}_{12},~\cdots,~{\bf p}_{1K}, ~{\bf p}_{21},~ \cdots,~ {\bf p}_{JK}],\\
{\bf H}_i &\!\!\!\!=\!\!\!\!& [{\bf h}_{i11},~ {\bf h}_{i12},~\cdots,~ {\bf h}_{i1K},~ {\bf h}_{i21},~ \cdots,~ {\bf h}_{iJK}]^T\!\!.\ \ \
\end{eqnarray}
\normalsize
In equations above, ${\bf P}$ is a $Q \times JK$ pilot code matrix, and ${\bf H}_i$ is a $JK \times M$ matrix. Note that if we treat all the $JK$ users in $J$ cells as if in one cell (with BS$_i$), then the system model in (\ref{received_signal_original}) becomes exactly the one in \cite{ICC2018}. In \cite{ICC2018}, we showed that BS$_i$ is able to estimate the large-scale channel gains of at most $N_{\max}=Q^2$ users if ${\bf P}$ is complex, and at most $N_{\max}=Q^2 - Q + 1$ users if ${\bf P}$'s entries have the same amplitude and only vary on phase. In this work, we assume $J \cdot K \leq N_{\max}$ so that we can treat these $J$ cells as a cluster/cell\footnote{In the context of a real network deployment involving a very large number of cells, $J$  can be viewed as  the number of cells that are in the vicinity of cell $i$. Effectively, users scheduled in cells outside this group of $J$ cells are sufficiently far away from cell $i$ so that their users' channel gains to cell $i$ are negligible.} and directly apply the large-scale channel-gain estimation scheme from \cite{ICC2018}.

For completeness, we briefly re-introduce the key idea of the scheme proposed in \cite{ICC2018} here. Essentially, we use the sample covariance matrix of the received signal at the BS to approximate its actual covariance matrix under the umbrella of massive MIMO and to exploit its full degrees of freedom. Specifically, each column of ${\bf Y}_i$ represents the signal vector (with $Q$ dimensions) seen at the corresponding antenna of the BS$_i$. Since each channel vector ${\bf h}_{ijk}$'s $M$ entries are i.i.d., the $m^{th}$ column of ${\bf Y}_i$, i.e., ${\bf y}_{im}$, is also i.i.d. over $m$. Investigation on ${\bf y}_{im}$ reveals that its covariance matrix can be written as:
\begin{equation}
{\bf R}_i = \mathbb{E}[{\bf y}_{im}{\bf y}_{im}^H]={\bf P}{\bf G}_i{\bf P}^H + \sigma^2_u {\bf I}_Q, \label{cov_to_ls}
\end{equation}
where ${\bf G}_i = \text{diag}([g_{i11}, ~g_{i12}, ~\dotsb, ~g_{ijk}, ~\dotsb, ~g_{iJK}])$ is a diagonal matrix. Since (\ref{cov_to_ls}) above consists of $Q^2$ linear equations in the unknown variables $g_{ijk}$'s, $Q^2$ unknowns can be resolved as long as the linear equations are linearly independent. However, ${\bf R}_i$ is not available in practice. Instead, we can use the sample covariance matrix of ${\bf Y}_i$, denoted by
\begin{eqnarray}
{\bf\hat{ R}}_i = \frac{1}{M} {\bf Y}_i{\bf Y}_i^H,
\end{eqnarray}
as an approximation of ${\bf R}_i$, as ${\bf\hat{ R}}_i$ converges to ${\bf R}_i$ when $M$ grows. Since ${\bf \hat{R}}_i \approx {\bf R}_i$, the estimation of ${\bf G}_i$, denoted by ${\bf\hat{G}}_i$, can be obatined by resolving the optimization problem:
\begin{subequations}\label{eqn:opt}
\begin{align}
{\bf\hat{G}}_i =  \arg \min_{{\bf \Theta}}
  & \;\;\; \|{\bf \hat{R}}_i-{\bf P}{\bf \Theta}{\bf P}^H - \sigma^2_u {\bf I}_Q\|_{\textrm F}^2 \\
 \textrm{s.t.} & \;\;\; {\bf \Theta}\succeq {\bf 0},
 \end{align}\label{large_est}
\end{subequations}
where ${\bf \Theta} = \text{diag}([\theta_{i11}, ~\theta_{i12}, ~\dotsb, ~\theta_{ijk}, ~\dotsb, ~\theta_{iJK}])$.

\subsection{Uplink: Small-scale Fading Estimation}

To perform MU-MIMO transmission to in-cell users, each BS needs to know the small-scale channel fading of the users whose large-scale gain has already been estimated. While a variety of estimators exist for resolving this problem, we consider the MMSE estimator for simplicity in this work. Since the estimation problem follows the standard form, we can directly obtain the estimation $\hat{{\bf H}}_i$ by treating the estimated $\hat{{\bf G}}_i$ as the actual ${\bf G}_i$:
\begin{equation}
\hat{{\bf H}}_i = \underbrace{\hat{{\bf G}}_i{\bf P}^H [{\bf P}\hat{{\bf G}}_i{\bf P}^H + \sigma^2_u {\bf I}_Q]^{-1}}_{\text{MMSE estimator}~{\bf D}}{\bf Y}_i.\label{MMSE_1}
\end{equation}
Observation of (\ref{MMSE_1}) reveals that the MMSE estimator ${\bf D}$ is a $JK\times Q$ matrix. Note that the BS has only $Q$ REs, i.e., $Q$-dimensional observations, to learn the small-scale fading. Thus, even though $\hat{{\bf H}}_i$ is an $JK \times M$ matrix, it has no more than $Q$ linear independent rows. This implies that when more than $Q$ nearby users (with large-scale channel gains significantly larger than zero) simultaneously send pilots to the BS, the BS can only resolve the small-scale channel fading of at most $Q$ users. To determine the users for which the BS continues to estimate the small-scale fading, we partition all the users into two groups namely ``group a" and ``group b". At BS$_i$, ``group a" consists of its $K$ intended users and $Q - K$ out-of-cell users with the {\em strongest} estimated large-scale fading (sorting the estimated large-scale fading of all out-of-cell users in a descending order and picking the first $K-Q$ users). Then the remaining $JK - Q$ out-of-cell users are assigned to ``group b". After rearranging (\ref{MMSE_1}), we obtain the complete channel estimation of the $Q$ users in ``group a" as
\begin{equation}
\hat{{\bf H}}_i^{\text{a}} = \hat{{\bf G}}_i^{\text{a}}{{\bf P}_i^{\text{a}}}^H [{\bf P}_i^{\text{a}}\hat{{\bf G}}_i^{\text{a}}{{\bf P}_i^{\text{a}}}^H + {\bf P}_i^{ \text{b}}\hat{{\bf G}}_i^{\text{b}}{{\bf P}_i^{\text{b}}}^H + \sigma^2_u {\bf I}_Q]^{-1}{\bf Y}_i,\label{MMSE_2}
\end{equation}
where ${(\cdot)}_i^{\text{a}}$, ${(\cdot)}_i^{\text{b}}$ denote parameters of interest in ``group a" and ``group b", respectively. Note that the small-scale fading of users in ``group b" are not of interest, as the BS cannot resolve more than $Q$ users' channels, whereas their large-scale fading estimations are still used to produce $\hat{{\bf H}}_i^{\text{a}}$.

Since the large-scale estimator in (\ref{eqn:opt}) is a least-squared estimator, the estimation performance of $\hat{g}_{ijk}$ depends on the real value $g_{ijk}$ and the equivalent noise term, which as mentioned in \cite{ICC2018} is associated back with $g_{ijk}$ (up to fourth-order statistics) and also how large $M$ is. Intuitively, $\hat{g}_{ijk}$ becomes less accurate when $g_{ijk}$ is close to or even below the noise power level. Thus, the large-scale channel estimation $\hat{{\bf G}}_i^{\text{b}}$ of the users in ``group b" is {\em not} as good as $\hat{{\bf G}}_i^{\text{a}}$ for users in ``group a", which could affect the quality of $\hat{{\bf H}}_i^{\text{a}}$.

To address the problem above, we propose a method of using a weight parameter $\beta \in [0,\  1]$ to tune the impact of $\hat{{\bf G}}_i^{\text{b}}$ on $\hat{{\bf H}}_i^{\text{a}}$, and the modified MMSE estimator is given by:
\begin{equation}
\hat{{\bf H}}_i^{\text{a}} = \hat{{\bf G}}_i^{\text{a}}{{\bf P}_i^{\text{a}}}^H [{\bf P}_i^{\text{a}}\hat{{\bf G}}_i^{\text{a}}{{\bf P}_i^{\text{a}}}^H + \beta{\bf P}_i^{ \text{b}}\hat{{\bf G}}_i^{\text{b}}{{\bf P}_i^{\text{b}}}^H + \sigma^2_u {\bf I}_Q]^{-1}{\bf Y}_i.\label{MMSE_3}
\end{equation}
In general, the value of $\beta$ can be optimized w.r.t. the system parameters such as cell size, path-loss factor, and user distribution. In this paper, we focus on only two extreme cases for simplicity: $\beta = 0$ and $\beta = 1$. While $\beta = 1$ implies that we assert $\hat{{\bf G}}_i^{\text{b}}$ is accurate enough, $\beta = 0$ indicates that $\hat{{\bf G}}_i^{\text{b}}$  is not satisfied at all. Thus, we would rather eliminate its impact and set it to be zero. In Sec. \ref{sec:sim}, we demonstrate the impact of $\beta$ on $\hat{{\bf H}}_i^{\text{a}}$ via numerical simulations.

\subsection{Downlink: Interference Mitigation and MU-MIMO}

After estimating the small-scale channel fading of users in ``group a" from (\ref{MMSE_3}), each BS performs ZFBF to multiplex their intended users in spatial domain. In particular, each BS first zero-forces the dimension along the channel vectors to the $Q-K$ out-of-cell users in ``group a", and then broadcasts the $K$ independent steams via MU-MIMO to the intended $K$ in-cell users in ``group a"\footnote{Due to the randomness of wireless channel and propagation environment, for each cell the number of in-cell users which also fall into ``group a" might be less than $K$. To improve the user rate performance, the BS can choose less than $K$ users to serve. Thus, it adds an additional variable for the optimization purpose, and this is out of the scope of this paper. On the other hand, under the umbrella of massive MIMO owing to channel hardening, the in-cell users are more likely to stay in ``group a" when $M$ grows.}. Specifically, each BS$_i$ follows the two steps below sequentially:
\begin{enumerate}
\item first computes the pseudo-inverse of $\hat{{\bf H}}_i^{\text{a}}$ (a $Q\times M$ matrix ) as $\hat{{\bf H}}_i^{\text{a}}{}^\dagger = \hat{{\bf H}}_i^{\text{a}}{}^H(\hat{{\bf H}}_i^{\text{a}}\hat{{\bf H}}_i^{\text{a}}{}^H)^{-1}$ (an $M\times Q$ matrix);

\item then normalizes each column of $\hat{{\bf H}}_i^{\text{a}}{}^\dagger$ to a unit vector:
\begin{eqnarray}
{\bf v}_{ik} = \hat{{\bf H}}_i^{\text{a}}{}^\dagger(:,k)/\|\hat{{\bf H}}_i^{\text{a}}{}^\dagger(:,k)\|.
\end{eqnarray}
The vectors ${\bf v}_{ik}$'s are the beamforming vectors for the $K$ intended users associated with BS$_i$.

\end{enumerate}

\subsection{Downlink: User Rate Performance}

The received signal at $\text{UE}_{ik}$ consists of three parts: desired signal, intra-cell interference and inter-cell interference. The signal-to-interference-plus-noise ratio (SINR) at $\text{UE}_{ik}$ can be readily written as
\begin{align*}
\text{SINR}_{ik} = \frac{|{\bf v}_{ik}^H {\bf h}_{iik}|^2}{\sum_{k' \neq k} |{\bf v}_{ik'}^H {\bf h}_{iik'}|^2 + \sum_{j \neq i} \sum_{k'} |{\bf v}_{jk}^H {\bf h}_{ijk'}|^2 + \sigma_d^2}.
\end{align*}
Besides intra-cell elimination via ZFBF, the inter-cell interference can also be suppressed through precoding, if the channel estimations are accurate.

\section{Novel Non-orthogonal Codebook Design}

In \cite{ICC2018} the Grassmannian-Line-Packing was considered as the non-orthogonal codebook design to estimate the large-scale fading channel gains of users. The large-scale channel-estimation performance can be evaluated via noise enhancement, which is characterized by ${\textrm{Tr}}(({\bf D}^H{\bf D})^{-1})$ (see Sec. IV in \cite{ICC2018}) where ${\bf D}$ is a $Q^2\times JK$ matrix given by:
\begin{eqnarray}
{\bf D} &\!\!\!\!=\!\!\!\!& [\textrm{vec}({\bf p}_{11}{\bf p}_{11}^H)~~ \textrm{vec}({\bf p}_{12}{\bf p}_{12}^H)~~\cdots ~~\textrm{vec}({\bf p}_{JK}{\bf p}_{JK}^H)]\ \ \ \ \\
&\!\!\!\!=\!\!\!\!&[{\bf p}_{11}^*\otimes {\bf p}_{11}~~{\bf p}_{12}^*\otimes {\bf p}_{12}~~\cdots ~~{\bf p}_{JK}^*\otimes {\bf p}_{JK}].\label{eqn:defineD}
\end{eqnarray}
In \cite{ICC2018}, we numerically showed that the Grassmannian-Line-Packing codebook is significantly better than the Gaussian codebook in terms of ${\textrm{Tr}}(({\bf D}^H{\bf D})^{-1})$ archiving a smaller value. On the other hand, while the Grassmannian codebook provides equally distinguishable pilot codes, it degrades the quality of small-scale estimation when we focus on local users in one cell, compared to that using orthogonal pilot codes. Specifically, in Grassmannian codebook design, the amplitude of the inner product between {\em any} two codes is minimized \cite{Hearth}, meaning that the angle between them is maximized. When we use such a codebook for large-scale channel estimation, the accumulative interference (from the other $JK-1$ users) projected to the one to be estimated would be minimized, and thus reducing the noise enhancement compared to Gaussian codebook while still guaranteeing $Q^2$ users can be resolved. However, for small-scale fading estimation, each BS only focuses on its up to $Q$ nearby/local users. For {\em only} those up to $Q$ local users, with the use of Grassmannian codebook, the BS still sees pilot contamination/ interference projected from the other $Q-1$ users. Since those $Q-1$ users are local, the interference would be relatively strong, and thus affecting the estimation quality. In contrast, when using orthogonal pilot codes for estimating the small-scale channel of those local $Q$ users, each user's channel estimation is free of the interference from the other $Q-1$ local users.

Now let us summarize the observations that we made. First, the Grassmannian codebook supports more than $Q$ and at most $Q^2$ users. It is preferable for the second-order channel statistics estimation of all $JK$ users, but not preferable for the first-order channel statistics estimation of the $Q$ local users. Second, the orthogonal codebook supports up to $Q$ users. It is preferable for the first-order channel statistics estimation of $Q$ local users, but subject to pilot contamination caused by pilot code reuse in neighbouring cells. While it is impossible for the Grassmannian codebook (with $JK>Q$) to have any subset of columns orthogonal to each other, we are interested in the question: {\em Is it possible to design a novel codebook to achieve a tradeoff between the Grassmannian codebook and the orthogonal codebook, so as to improve the overall throughput/rate performance?}

If such a new codebook can be designed, intuitively it should have the two properties below:
\begin{enumerate}
\item[P1.]  The pilot codes/vectors allocated to the $K$ users in the same cell can be distinguished from each other as much as possible. Ideally, they are orthogonal to each other.
\item[P2.] Under the premise of satisfying P1, the users in the neighbouring cells can also be distinguished from each other by the anchor BS as much as possible. That is, the resulting ${\textrm{Tr}}(({\bf D}^H{\bf D})^{-1})$ should be small. Ideally, the maximum of the inner-product amplitude between any two pilot vectors assigned to different cells is minimized, just like the Grassmannian-Line-Packing design.
\end{enumerate}

In this section, we answer to the question above with the positive. Specifically, we provide a new scheme for codebook design by borrowing the concept of mutually unbiased basis (MUB) which has been investigated and is still a key topic in quantum information theory \cite{MUB_1, MUB_revisit, MUB_construct, real_MUB, MUB_2_5}.

\subsection{The Problem Formulation of New Code Design}

For the $JK$ users in the system, we need to design $J$ matrices of ${\bf P}_1, \cdots, {\bf P}_J$ where each ${\bf P}_j$ is a $Q\times K$ matrix and its $k^{th}$ column is the pilot code assigned to UE$_{jk}$. Since the new codebook should have the two properties above, we use the following two conditions to characterize the two properties:
\begin{enumerate}
\item[C1.]{${\bf P}_j^H {\bf P}_j = {\bf I}_K$, for $\forall j\in \mathcal{J}\triangleq \{1,\cdots,J\}$.}
\item[C2.]{$\min_{\{{\bf P}_j,j\in\mathcal{J}\}}\max_{\{k_1,k_2\}}|({\bf P}_i^H {\bf P}_j)_{k_1,k_2}| = c$, for $\forall j \neq i$, $i,j\in\mathcal{J}$, and $\forall~k_1, k_2 \in\{1,\cdots,K\}$, where $c$ is a constant and should be as smaller as possible. Ideally, we desire $|({\bf P}_i^H {\bf P}_j)_{k_1,k_2}| = c$.}
\end{enumerate}
For brevity of presentation, we denote by ${\bf P}(J, Q, K)$ a set of matrices ${\bf P}_1, \cdots, {\bf P}_J$ that satisfies both C1 and C2 above. In addition, we use $\{{\bf P}(J, Q, K)\}$ to denote the collection of all ${\bf P}(J, Q, K)$ given the parameters $(J, Q, K)$.

In the conditions above, C1 indicates that the pilot codes assigned to the users in the same cell are orthogonal among themselves, which inherits the benefits of orthogonal codebooks; C2 indicates our desire to inherit the attributes of Grassmannian codebook, and the value of the constant $c$ is to-be-optimized, which in general is an open problem but we shed light on this problem in Sec. IV.E.

To design a codebook ${\bf P}$ satisfying C1 only is not difficult, as one can arbitrarily choose $J$ unitary matrixes and pick the first $K$ columns to each unitary matrix form each ${\bf P}_j$. However, how to jointly design ${\bf P}_j$'s to satisfy C2 as well is not straightforward. Indeed, even the existence of such a codebook is not apparent. Fortunately, the conditions of C1 and C2 are strongly associated with an extensively studied topic in quantum information theory, called mutually unbiased bases (MUB). We thereby can leverage available MUB results for our codebook design, as introduced in next section.

\subsection{Mutually Unbiased Bases (MUB)}

\begin{definition} \cite{MUB_1} {\em (MUB)}
Let ${\bf X}=\{{\bf x}_1,... , {\bf x}_Q\}$ and ${\bf Y}=\{{\bf y}_1,... , {\bf y}_Q\}$ be two distinct orthogonal bases in $\mathbb{C}^Q$. They are called unbiased bases if
\small
\begin{equation}
\frac{|{\bf x}^H_{q_1} {\bf y}_{q_2} |}{\| {\bf x}_{q_1} \|  \cdot\| {\bf y}_{q_2} \|} = \frac{1}{\sqrt{Q}} \quad \text{for all ${q_1},{q_2}\in\{1,\cdots,Q\}$.} \label{MUB_define}
\end{equation}
\normalsize
A set of orthogonal bases is a set of MUB if all pairs of distinct bases are unbiased.
\end{definition}

We denote by MUB$(J, Q, Q)$ a set of $J$ square matrices, each with $Q\times Q$ that form a set of MUB defined above, where the resulting constant $c = \frac{1}{\sqrt{Q}}$. It can be seen that both C1 and C2 are satisfied, and thus MUB$(J, Q, Q)$ can be directly used as a codebook ${\bf P}(J, Q, Q)$, where each sub-matrix ${\bf P}_j$ is a $Q \times Q$ square matrix. Next, we summarize some MUB-related results available so far, which form the basis for our new codebook design in this paper.
\begin{corollary} \label{corol_1}
\cite{MUB_revisit} MUB$(J=3, Q, Q)$ exists for any $Q \geq 2$.
\end{corollary}
\begin{corollary} \label{corol_2}
\cite{MUB_1} MUB$(J,Q,Q)$ does not exist when $J>Q+1$ for any $Q \geq 2 $.
\end{corollary}

{\em Remark:} Corollary \ref{corol_1} provides a lower bound $J\geq 3$ for any $Q\geq 2$, and MUB$(3, Q, Q)$ can be constructed via the approach introduced in \cite{MUB_revisit}. In turn Corollary \ref{corol_2} provides an upper bound $J\leq Q+1$. So far, it is known that this upper bound is tight in dimensions $Q$ which is a power of a prime \cite{MUB_construct}. A simple approach to construct all MUB$(Q + 1, Q, Q)$ for prime $Q$'s was provided in \cite{MUB_revisit}, and examples of MUB$(Q + 1, Q, Q)$ for $Q = 2, 3, 4$ were provided in \cite{MUB_1}. However, when $Q$ is {\em not} a power of a prime, e.g., $Q = 6$, whether this upper bound is tight remains open.

Besides the complex-valued matrices construction, in practical wireless system codebook design, we are also interested in the cases including restricting all the entries of ${\bf P}_j$'s to be real-valued, or to have a constant amplitude but phases can vary. We denote the MUBs satisfying these constrains as MUB$^r(J,Q,Q)$ and  MUB$^{\phi}(J,Q,Q)$, respectively. With these constraints, we summarize some useful results from prior works as follows:
\begin{corollary}
\cite{real_MUB} MUB$^r(J,Q,Q)$ does not exist when $J>Q/2 + 1$ for any $Q \geq 2$.\label{corol_3}
\end{corollary}

{\em Remark:} The upper bound $J \leq Q/2 + 1$ compared to that in Corollary 2 is reduced due to losing variety along the entry-amplitude dimension. Generally, this bound is still loose, and some tighter bounds are derived and tabulated for some specific $Q$, e.g., in \cite{real_MUB}.

\begin{corollary}
\cite{MUB_2_5} Given MUB$(J,Q,Q)$ for any $Q \geq 2 $, MUB$^{\phi}(J-1,Q,Q)$ exists as well and can be constructed from MUB$(J,Q,Q)$. \label{corol_4}
\end{corollary}

{\em Remark:} As shown in \cite{MUB_2_5}, once MUB$(J,Q,Q)$ are available, MUB$^{\phi}(J-1,Q,Q)$ can be easily constructed, and vice versa. The approach is illustrated via an example in the next section. Note that for Grassmannian-Line-Packing design, such a construction approach does not exist.

\subsection{The Optimality of MUB Codebooks on Noise Enhancement}

In Sec. IV.A, we explained that designing a codebook satisfying both C1 and C2 is much more difficult than that satisfying C1 only. Suppose ${\bf \tilde{P}}(J, Q, K)$ denotes a set of matrices ${\bf P}_1, \cdots, {\bf P}_J$ that satisfies C1 only, and $\{{\bf \tilde{P}}(J, Q, K)\}$ is the collection of all ${\bf \tilde{P}}(J, Q, K)$. Moreover, in Sec. IV.B, we show that MUB satisfy both C1 and C2. Apparently, ${\textrm{MUB}}(J, Q, Q) \subset{\bf P}(J, Q, K) \subset {\bf \tilde{P}}(J, Q, K)$.

Recall that $\Tr(({\bf D}^H{\bf D})^{-1})$ measures the noise enhancement in large-scale channel estimation. To minimize it, we have the following result:
\begin{theorem}
For any ${\bf P} \in \{{\bf \tilde{P}}(J, Q, Q)\}$, $\Tr(({\bf D}^H{\bf D})^{-1})$ is minimized when ${\bf P} = {\textrm{MUB}}(J, Q, Q)$.\label{theorem_1}
\end{theorem}

{\it Proof:} The proof is provided in Appendix \ref{app-2}.
\hfill\QED

{\it Remark:} This theorem implies that under the premise of satisfying C1, MUB$(J,Q,Q)$ is the best code for $K=Q$ in terms of the noise enhancement performance. Clearly, each unitary matrix ${\bf P}_j$ inherits the attributes of orthogonal codebooks and the definition of MUB keeps the attributes of Grassmannian codebooks.

\subsection{An Example of Applying MUB for Codebook Design}

For the reader to better understand how to use MUB as codebook in a multi-cell system, we consider a simple system with $J=2$  cells, each with $Q = 2$ REs serving $K=2$ users simultaneously. Consider the following 2 matrices in $\mathbb{C}^{2\times 2}$:
\begin{equation*}
{\bf P}_1 = \frac{1}{\sqrt{2}} \begin{bmatrix} 1 & 1\\ 1 & -1\\ \end{bmatrix},~~
{\bf P}_2 = \frac{1}{\sqrt{2}} \begin{bmatrix} 1 & 1\\ i & -i\\ \end{bmatrix}.
\end{equation*}
It can be easily verified that the above two matrices form a set of MUB$^{\phi}(2,2,2)$, because all the entries of both matrices have the same amplitude, and they satisfy both C1 and C2 conditions. Each BS picks one matrix above exclusively as its own codebook, say, BS$_1$ chooses ${\bf P}_1$ and BS$_2$ chooses ${\bf P}_2$, and both BSs completely know ${\bf P} = [{\bf P}_1, {\bf P}_2]$. By following the transmission scheme proposed in Sec. III, each BS is able to estimate the large-scale fading of all the $4$ users and the small-scale fading of its $2$ associated users. Moreover, if only one user is active in each cell, then the BS can use the one extra dimension to mitigate the interference towards out-of-cell users, thus reducing the overall inter-cell interference in the system. 
Note that in this case, {\em no matter} which user is activated, we do not have the pilot contamination caused by the same pilot reuse over different cells, as each of the 4 pilot vectors is unique.

Furthermore, based on Corollary \ref{corol_4}, MUB$(3,2,2)$ can be easily constructed from MUB$^{\phi}(2,2,2)$. To see this, we can arbitrarily pick a $Q\times Q$ unitary matrix ${\bf U}$ and form another two matrices ${\bf U}{\bf P}_1$ and ${\bf U}{\bf P}_2$. It can be easily verified that $[{\bf U}, {\bf U}{\bf P}_1, {\bf U}{\bf P}_2]$ is an MUB$^{\phi}(3, 2, 2)$. Generally, given a set of MUB$^{\phi}(J ,Q, Q) $$=  [{\bf P}_1, \cdots, {\bf P}_J]$, using any $Q \times Q$ unitary matrix ${\bf U}$, we can obtain MUB$(J+1 ,Q, Q)$ $=  [{\bf U}, {\bf U}{\bf P}_1, \cdots, {\bf U}{\bf P}_J]$.

\subsection{Overview of the More General Codebook Design}

Clearly, MUB$(J, Q, Q)$ is a special case of ${\bf P}(J,Q,K)$. If $K=Q$, ${\bf P}(J,Q,K)$ becomes MUB$(J, Q, Q)$, and the upper bound $J\leq Q+1$ from Corollary \ref{corol_2} also applies here. Moreover, if $K=1$, ${\bf P}(J,Q,1)$ becomes Grassmannian-Line-Packing codes, and thus the upper bound $J\leq Q^2$ directly applies as well \cite{ICC2018}. Other than these two extreme cases, it is natural to ask: {\em Does ${\bf P}(J, Q, K)$ exist when $1<K<Q$? If yes, how to construct it and what values does $J$ can take?} To the best of our knowledge, no results have been published on such problems. Nevertheless, based on the observations and intuitions so far, we could take a further outlook on these open problems, as the solution to these problems might provide more flexility in new codebook design in future. In the following, we directly state our results:
\begin{proposition}
For ${\bf P}(J, Q, K)$ where $K = 2, \dotsb,  Q-1$, the upper bound of $J$ ranges from $Q+1$ to $Q^2$, inclusively.
\end{proposition}

{\it Proof}: Given a set of MUB$(Q+1, Q, Q)$, a general ${\bf P}(Q+1, Q, K)$ can be constructed by retaining only the first $K$ column of each ${\bf P}_j$. In addition, if $J > Q^2$ for ${\bf P}(J, Q, K)$ and some $K$, then it conflicts with \cite{Hearth} (see its Theorem 2.3 and Corollary 2.4) that at most $Q^2$ Grassmannian vectors exist in $\mathbb{C}^Q$. Hence, if ${\bf P}(J, Q, K)$ exists, $J$ must be upper bounded by some integer between $Q+1$ and $Q^2$, inclusively.

\begin{proposition}
If ${\bf P}(J, Q, K)$ exists, the following lower bound on the constant $c$ (see the condition C2) always holds:
\small
\begin{equation}
c \geq \sqrt{\frac{JK - Q}{KQ(J-1)}} \label{bound}
\end{equation}
\normalsize
\end{proposition}

{\it Proof}: This generalizes the Welch bound shown in \cite{Hearth}. Please see Appendix \ref{app-1} in detail.
\hfill\QED

{\it Remark:} As shown in \cite{ICC2018}, the parameter $c$ characterizes the noise enhancement in large-scale channel gains estimation, and the lower the better. The lower bound provided in (\ref{bound}) points out the best we can do. Note that (\ref{bound}) is tight when $K=1, Q$, because it reduces to $c \geq \sqrt{\frac{J - Q}{Q(J-1)}}$ for Grassmannian line packing ($K=1$), and to $c \geq 1/\sqrt{Q}$ for MUB ($K=Q$).

\section{Simulations}\label{sec:sim}

In this section, we evaluate the performance of the scheme that we proposed in Sec. III by using the novel codebook design we introduced in Sec. IV via simulations.

\begin{figure}[!t]
\centering
\includegraphics[width = 0.25\textwidth]{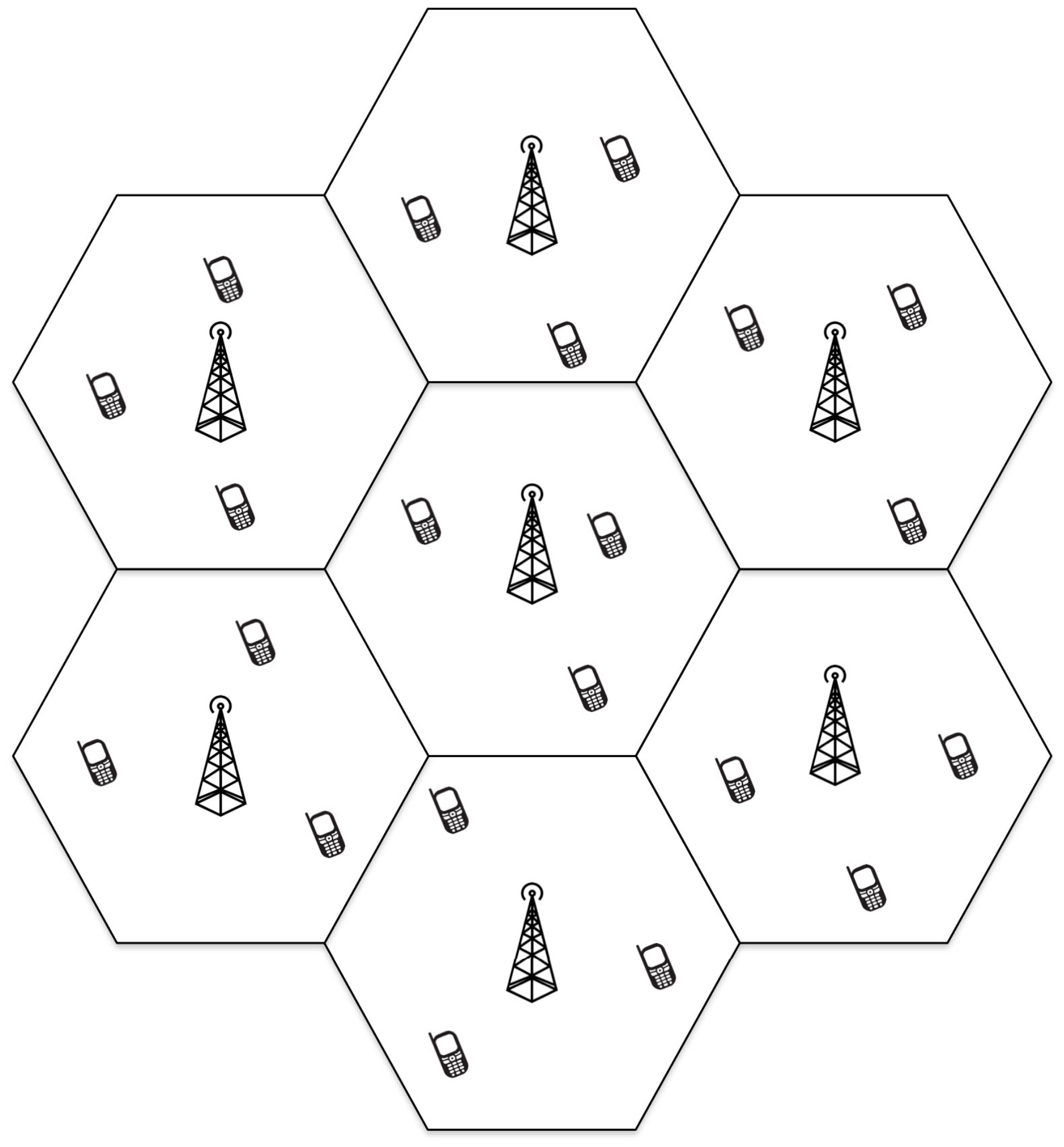}
\caption{System model used for simulation}
\label{fig_model}
\end{figure}

We consider a system with $7$-cell ($J=7$) wrap-around hexagon grid as shown in Fig. \ref{fig_model}, where each BS is equipped with $M = 128$ antennas and serves $K$ ($K \leq Q$) single-antenna UEs per RB simultaneously. Each BS uses $Q = 7$ REs per RB for uplink training. While the BS is assumed to be located at the cell center, its $K$ associated users are uniformly distributed within its cell. We assume that the radius of each cell is $R=10$ m, the path-loss factor is $\alpha=2.5$, and the large-scale channel gain between $\text{UE}_{jk}$ and $\text{BS}_i$ is $g_{ijk} = d_{ijk}^{-\alpha}$ where $d_{ijk}$ is the distance between them. Moreover, we assume that the uplink effective noise power is $\sigma^2_u = 10^{-3}$ (i.e., the uplink transmit SNR$=30$ dB), and the downlink effective noise power is $\sigma^2_d = 10^{-4}$ (i.e., the downlink transmit SNR$=40$ dB). Our goal is to investigate the effective user rate for each UE$_{ik}$, which is defined as the nominal/peak user rate $\log_2(1 + \textrm{SINR}_{ik})$ scaled by the factor $K/Q$ indicating the user activity fraction, where $\textrm{SINR}_{ik}$ is defined in Sec. III.D.

In this section, we use $(J,Q,Q)$ as the parameters for codebook dimensions, where each cell is assigned a $Q\times Q$ pilot matrix (i.e., $Q$ pilot code vectors are available for use in each cell). Note that because of $K\leq Q$, the users in each cell might not use up all the pilot codes assigned to their cell. In the following, we consider the use of the 4 types of codebooks. Note that we directly set $(J,Q,Q)=(7,7,7)$, and since each BS serves only $K$ user per RB in each cell, in each iteration, each user in every cell is randomly assigned a pilot code from the $7$ available codes exclusively.
\begin{enumerate}
\item MUB$^{\phi}(7, 7, 7)$: We directly use MUB$^{\phi}(7, 7, 7)$ which is constructed by using the approach provided in \cite[Sec. 2.4]{MUB_revisit} as the codebook ${\bf P}(7, 7, 7)$.

\item MUB$(7,7,7)$: We first construct MUB$(8,7,7)$ by augmenting MUB$^{\phi}(7, 7, 7)$ with a randomly generated $7 \times 7$ unitary matrix. Then we uniformly select $7$ matrixes out of MUB$(8,7,7)$ to form the MUB$(7,7,7)$.

\item \textit{Incomplete codebook} ${\bf \tilde{P}}(7, 7, 7)$: To make a benchmark for MUB-based codes, we construct it by stacking up seven randomly generated $7\times 7$ unitary matrices, satisfying the constrain C1 only.

\item \textit{$7 \times 7$ Orthogonal Codebook:} We also consider the performance of the orthogonal codebook, as a benchmark for our proposed non-orthogonal training scheme. In particular, the {\em same} randomly generated $7 \times 7$ orthogonal codebook is reused in each cell. During uplink training, each user is randomly assigned a pilot code exclusively from its cell, and each BS estimates users' channels by using zero-forcing beamforming. During downlink transmission, each BS transmits the data stream to intended users via ZFBF. Moreover, we assume that all the BSs are only aware of their own users\footnote{Although the same codebook is used in each cell, when $K<Q$, {\em any specific} pilot code reuse factor is actually larger than one, as any two users from different cells might select different pilot codes from the same codebook. Thus, this approach outperforms traditional operation using orthogonal codebooks in terms of user rates, and hence serving as an aggressive benchmark.}.
\end{enumerate}

\begin{figure}[!t]
\centering
\includegraphics[width = 0.48\textwidth]{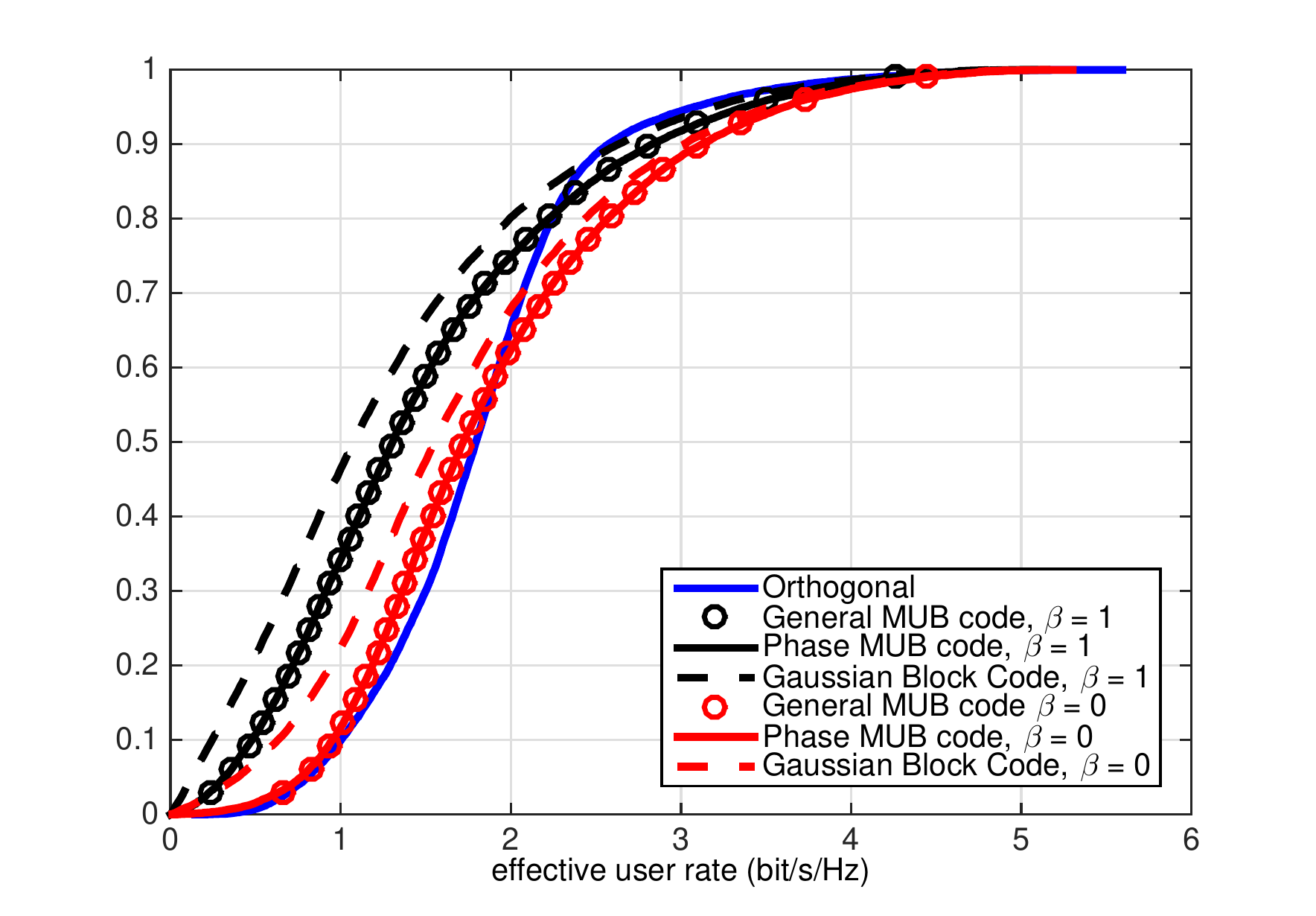}\vspace{-0.15in}
\caption{Distribution of effective user rates for $K = 3$ considering a variety of codebook design and small-scale channel-fading estimator design}
\label{fig_1}
\end{figure}

In Fig. \ref{fig_1} we show the CDF (Cumulative Distribution Function) curves of the user rates for $K=3$. It can be seen that the performance of $\beta = 0$ (red-colored) is much better than that of $\beta = 1$ (black-colored) under the current system configuration. As mentioned in Sec. III, it indicates that $\beta$ affects the performance and thus needs carefully tuning depending on the practical circumstance. Next, it can be seen that MUB$(7,7,7)$ and MUB$^{\phi}(7,7,7)$ both have almost the same performance and they are both significantly better than ${\bf \hat{P}}(7,7,7)$. Finally, comparing the performance of MUB$(7,7,7)$ to that of the orthogonal training scheme, one can see that MUB$(7,7,7)$ is almost as good as orthogonal training in the low SINR regime, and apparently much better in the high SINR regime.

Based on the analysis so far, there are two benefits of using non-orthogonal uplink pilot training: (i) As the BS is aware of the channels of nearby out-of-cell users (i.e., cell-edge users in the neighboring cell), it can mitigate the interference projected to those users via ZFBF, and thus improving the achievable rates of those cell-edge users. (ii) As shown in \cite{ICC2018}, each BS is able to estimate the large-scale channel-gains of $Q^2$ users, which provides additional information compared to orthogonal training when we estimate the small-scale channel fading.

\begin{figure}[!t]\vspace{-0.15in}
\centering
\includegraphics[width = 0.48\textwidth]{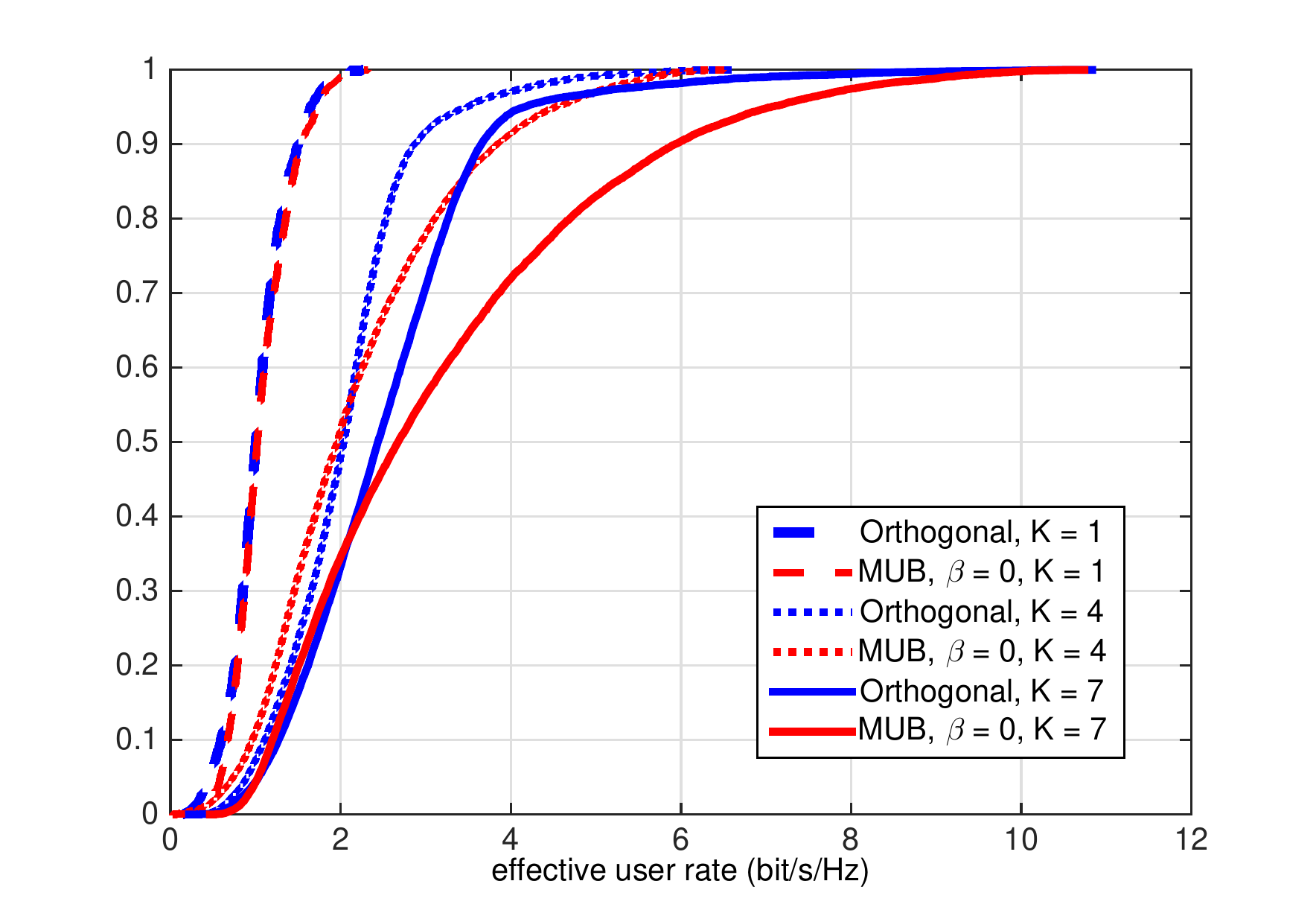}\vspace{-0.15in}
\caption{Distribution of effective user rates for MUB-based codes and orthogonal codes $K = 1,4,7$}
\label{fig_2}
\end{figure}

Next, we look into the CDFs of effective user rates by fixing $\beta=0$ but setting $K=1,4,7$. As Fig. \ref{fig_2} demonstrates, the advantage of non-orthogonal codes is more prominent as $K$ increases. In contrast, when using orthogonal training£¬pilot collision exists for all users if $K=7$, and all downlink transmissions generate direct interference to all the other users who use the same pilot code for uplink training. 
As per non-orthogonal training, even the BS does not zero force the interference to nearby out-of-cell users, it can still estimate the large-scale channel gains of its in-cell users more accurately than using orthogonal codes.

\begin{figure}[!t]\vspace{-0.15in}
\centering
\includegraphics[width = 0.48\textwidth]{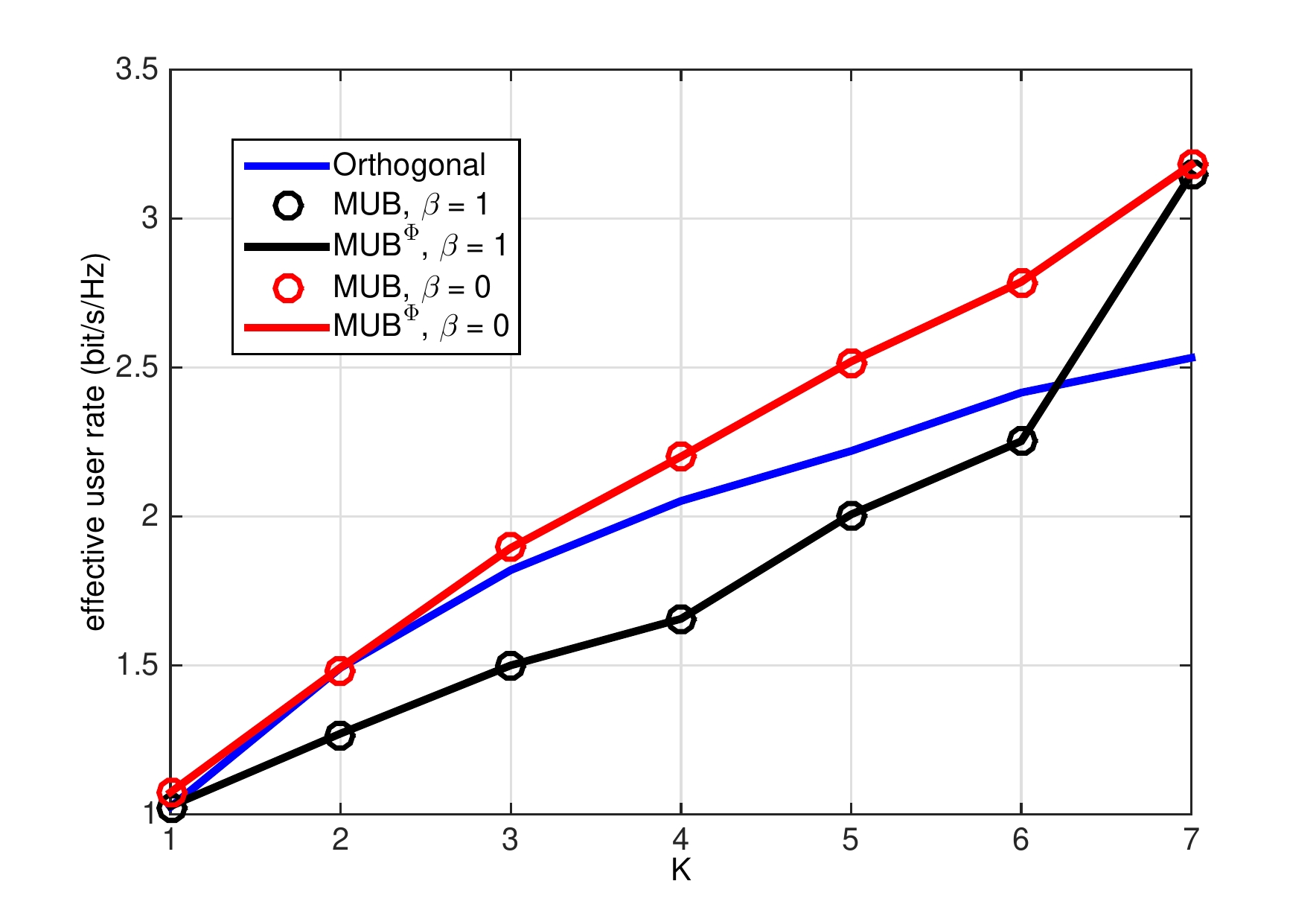}\vspace{-0.15in}
\caption{Average of effective user rates against $K$}
\label{fig_3}
\end{figure}

Finally, we compare among the effective user rates of MUB$(J,Q,Q)$ and MUB$^{\phi}(J,Q,Q)$ codes with $\beta=0,1$ against the number of active user $K$, as shown in Fig. \ref{fig_3}. It can be seen that the performance by using non-orthogonal codes with $\beta=0$ is always better than that using orthogonal codes. Also, the relative rate gain (i.e., the gap) between them increases as $K$ grows. Another observation is that as $K$ increases, the gap between using $\beta = 0$ and $\beta = 1$ is diminished. This is due to the fact that the intended users in ``group a" also increases users when $K$ grows. Since the entries of $\hat{{\bf G}}_i^{\text{a}}$ are larger than that of $\hat{{\bf G}}_i^{\text{b}}$, $\hat{{\bf G}}_i^{\text{a}}$ becomes dominant in (\ref{MMSE_3}) and the impact of $\beta$ is marginalized.

\section{Conclusion}

A class of pilot code designs for TDD-based uplink training in a massive MIMO cellular network is investigated in this paper. Using the novel channel estimation scheme with non-orthogonal codebook design and with interference mitigation proposed in this paper, we demonstrate user-rate improvement with respect to conventional operation using orthogonal codebooks. Several interesting problems could be investigated in future works, including investigating a number of theoretical open problems on the properties of the novel codes that we raised in Sec. IV.E, as well as simulation with more realistic channel models.

\appendix

\subsection{The Proof of Theorem 1}\label{app-2}

To make the proof easier to read, we first list the notations that we use in this section. In particular, we use ${\bf D}={\bf D}({\bf P})$ as the matrix define in (\ref{eqn:defineD}) by using the matrix ${\bf P}$ where ${\bf P} = $ MUB$(J, Q, Q)$. Similarly, we use $\tilde{{\bf D}}=\tilde{{\bf D}}({\tilde{\bf P}})$ as the matrix define in (\ref{eqn:defineD}) by using the matrix $\tilde{{\bf P}}$, where $\tilde{{\bf P}}$ is any codebook in $\{{\bf \tilde{P}}(J, Q, Q)\}$. Therefore, to prove Theorem \ref{theorem_1}, it suffices to show $\Tr(({\bf D}^H{\bf D})^{-1}) \leq \Tr((\tilde{{\bf D}}^H\tilde{{\bf D}})^{-1})$. Since the proof is sophisticated, the entire proof is broken down into several lemmas sequentially.

\begin{lemma} \label{lemma_1}
The {\em non-zero} eigenvalues of ${\bf D}^H{\bf D}$ are given by
\begin{equation}
  \lambda_k = \begin{cases}
    J, & k = 1,\\
    1, & k = 2, \cdots, JQ - J + 1.
  \end{cases}\label{eigenvalue}
\end{equation}
The eigenvector associated with $\lambda_1=J$ is $[1, 1, \cdots, 1]^T$, and the $JQ-J$ eigenvectors associated with the eigenvalue $1$ are $[\cdots,0,-1,0,\cdots,0,1,0,\cdots]^T$, which is a $JQ\times 1$ vector of zeros except for two non-zero entries where the $m$-th entry is $1$ and the $n$-th entry is $-1$, and where $m$ sequentially takes $1, (Q+1), \cdots, (JQ - Q + 1)$ and $n$ sequentially takes $(m + 1), (m+2), \cdots, (m + Q - 1)$, respectively.
\end{lemma}

{\em Proof:} Because of ${\bf P} = $ MUB$(J, Q, Q)$, we have:
\begin{equation}
\begin{cases}
    {\bf P}_j^H{\bf P}_j = {\bf P}_j{\bf P}_j^H = {\bf I}_Q,& \forall~j \in \mathcal{J}\\
    |{\bf p}_{iq_1}^H{\bf p}_{jq_2}| = \frac{1}{\sqrt{Q}} \triangleq c, &{\textrm{for}}~\forall~i\neq j,~{\textrm{and}}~\forall~q_1,q_2.
  \end{cases}
\end{equation}
Owing to Proposition 1 in \cite{ICC2018}, ${\bf D}^H{\bf D}$ can be expressed as
\begin{align}
 {\bf D}^H{\bf D} = \left[\begin{array}{rrrr}
   {\bf I}_Q & c^2 \cdot {\bf 1}_Q & \dotsb & c^2 \cdot {\bf 1}_Q\\
   c^2 \cdot {\bf 1}_Q & {\bf I}_Q & \dotsb & c^2 \cdot {\bf 1}_Q\\
   \vdots & \vdots & \ddots & \vdots\\
   c^2 \cdot {\bf 1}_Q & c^2 \cdot {\bf 1}_Q & \dotsb & {\bf I}_Q\\
   \end{array}\right]_{JQ \times JQ}. \label{temp_563}
\end{align}
Due to the special structure of ${\bf D}^H{\bf D}$, it can be easily verified that its eigenvalues and eigenvectors are those stated in this Lemma.
\hfill\QED

\begin{lemma}
The sum of the $Q$ columns of $\tilde{{\bf D}}_j=\tilde{{\bf D}}_j(\tilde{{\bf P}}_j)$ is a constant vector and identical for very $j \in \mathcal{J}$. In particular, $\sum_{q =1}^Q \tilde{{\bf d}}_{jq} = \tilde{{\bf d}}_{\textrm{s}}$, which is a $Q^2\times 1$ vector of zero except for the $(Q(q-1)+q)$-th entry equal to 1 where $q=1,\cdots,Q$. \label{lemma_2}
\end{lemma}

{\em Proof:}
Based on the definition in (\ref{eqn:defineD}), $\tilde{{\bf D}}_j(:,q)$ can be written as $\tilde{{\bf d}}_{jq} = \tilde{{\bf p}}_{jq}^* \otimes \tilde{{\bf p}}_{jq}$. Also, due to ${\bf \tilde{P}} \in \{{\bf \tilde{P}}(J, Q, Q)\}$ satisfying the condition C1, we have:
\begin{equation}
{\bf \tilde{P}}_j^H{\bf \tilde{P}}_j = {\bf  \tilde{P}}_j{\bf \tilde{P}}_j^H = {\bf I}_Q, ~~~~ \forall~j \in\mathcal{J}, \label{temp_582}
\end{equation}
where ${\bf  \tilde{P}}_j{\bf \tilde{P}}_j^H = {\bf I}_Q$ can be rewritten in a vector-form as
\begin{eqnarray}
\tilde{{\bf P}}_j(q_1,:) (\tilde{{\bf P}}_j(q_2,:))^H=
\begin{cases}
1, ~~~~q_1=q_2\\
0, ~~~~q_1\neq q_2.\label{temp_593}
\end{cases}
\end{eqnarray}
Recall again the definition in (\ref{eqn:defineD}), $\tilde{{\bf d}}_{jq}=\tilde{{\bf p}}_{jq}^*\otimes \tilde{{\bf p}}_{jq}$ is an $Q^2 \times 1$ vector. Hence, the $m$-th entry of $\tilde{{\bf d}}_{jq}$ is given by:
\begin{equation}
\tilde{{\bf d}}_{jq}(m) = \tilde{{\bf p}}_{jq}^*(q_1)\cdot \tilde{{\bf p}}_{jq}(q_2),~~~~m=Q(q_1-1)+q_2\label{temp_597}
\end{equation}
where $q_1,q_2$ both range over $1,\cdots,Q$.

Next, consider the $m$-th entry of $\tilde{{\bf d}}_{\textrm{s}}$ as follows:
\begin{align}
\tilde{{\bf d}}_{s}(m) &= \sum_{q=1}^Q \tilde{{\bf d}}_{jq}(m) = \sum_{q=1}^Q \tilde{{\bf p}}_{jq}^*(q_1)\cdot \tilde{{\bf p}}_{jq}(q_2)\label{eqn:dp}\\
& =(\tilde{{\bf P}}_j(q_1,:))^* (\tilde{{\bf P}}_j(q_2,:))^T\\
& =(\tilde{{\bf P}}_j(q_1,:) \tilde{{\bf P}}_j(q_2,:))^H)^* = \begin{cases}
   1, & q_1=q_2\\
   0, & q_1\neq q_2
  \end{cases}\label{eqn:dsum}
\end{align}
where (\ref{eqn:dp}) is obtained due to (\ref{temp_597}), and (\ref{eqn:dsum}) follows frrom (\ref{temp_593}). So far, since $\tilde{{\bf d}}_{\textrm{s}}(m)=1$ if $m\in\{(Q(q-1)+q)|q=1,\cdots,Q\}$, and $\tilde{{\bf d}}_{\textrm{s}}(m)=0$ for otherwise, we completes the proof.
\hfill\QED

\begin{lemma}
rank($\tilde{{\bf D}}^H\tilde{{\bf D}}$) $= JQ - J + 1$ almost surely.
\end{lemma}

{\em Proof:} Based on Lemma \ref{lemma_2}, since the sum of the $Q$ columns of each $\tilde{{\bf D}}_j$ are equal, 
it implies that the column space of $\tilde{{\bf D}}_1$ has at least one-dimensional intersection with the column space of each $\tilde{{\bf D}}_j$ for $j=2,\cdots,J$.
Therefore, when we form the matrix $\tilde{{\bf D}}_j=[\tilde{{\bf D}}_1,\cdots,\tilde{{\bf D}}_J]$, its column rank will be reduced from $JQ$ by at least $J-1$. Thus,
\begin{eqnarray}
{\textrm{rank}}(\tilde{{\bf D}}^H\tilde{{\bf D}})\leq {\textrm{rank}}(\tilde{{\bf D}}) \leq JQ - J +1.\label{eqn:lemma3_upper}
\end{eqnarray}
On the other hand, Lemma 1 already shows a specific example of $\tilde{{\bf D}}^H\tilde{{\bf D}}$ where ${\textrm{rank}}(\tilde{{\bf D}}^H\tilde{{\bf D}})= JQ - J +1$. Under the sense of almost surely, it implies that ${\textrm{rank}}(\tilde{{\bf D}}^H\tilde{{\bf D}})$ is at least $JQ - J +1$. This is because the determinant of ${\textrm{rank}}(\tilde{{\bf D}}^H\tilde{{\bf D}})$ is a $JQ$-order polynomial, whose non-zero eigenvalues either do not exist or constitute a subset with Lebesgue measure 1. Since Lemma 1 already specifies $JQ-J+1$ non-zero eigenvalues of $\tilde{{\bf D}}^H\tilde{{\bf D}}$, under the sense of almost surely, we have:
\begin{eqnarray}
{\textrm{rank}}(\tilde{{\bf D}}^H\tilde{{\bf D}})\geq JQ - J +1.\label{eqn:lemma3_inner}
\end{eqnarray}
Combining (\ref{eqn:lemma3_upper}) and (\ref{eqn:lemma3_inner}) results in rank($\tilde{{\bf D}}^H\tilde{{\bf D}}$) $= JQ - J + 1$ almost surely.
\hfill\QED

\begin{lemma}
$\sum_m\sum_n (\tilde{{\bf D}}^H\tilde{{\bf D}})_{m,n} \geq \sum_m\sum_n  ({\bf D}^H{\bf D})_{m,n}$
\end{lemma}

{\em Proof:} As derived later in (\ref{F_norm}) and (\ref{lower_bound2}) in Appendix B, $\sum_m\sum_n (\tilde{{\bf D}}^H\tilde{{\bf D}})_{m,n} \geq J^2Q$. In addition, owing to (\ref{temp_563}), $\sum_m\sum_n  ({\bf D}^H{\bf D})_{m,n} = J^2Q$, thus we complete the proof.
\hfill\QED

Next, suppose the largest eigenvalue of $\tilde{{\bf D}}^H\tilde{{\bf D}}$ is denoted by $\tilde{\lambda}_1$. In addition, because of Lemma 1 and Lemma 3, $\tilde{{\bf D}}^H\tilde{{\bf D}}$ and ${\bf D}^H{\bf D}$ both have rank $JQ-J+1\triangleq R$. Hence, we denote the $R$ non-zero eigenvalues of $\tilde{{\bf D}}^H\tilde{{\bf D}}$ and ${\bf D}^H{\bf D}$ by $\{\tilde{\lambda}\} = \{\tilde{\lambda}_1, \tilde{\lambda}_2, \cdots, \tilde{\lambda}_R\}$ and $\{\lambda\} = \{ \lambda_1, \lambda_2, \cdots, \lambda_R\}$ respectively, where the $R$ eigenvalues are sorted in a descending order. Also, we have the following definition:
\begin{definition}
For two equally-sized lists $\{x\}$ and $\{y\}$, we use ``$\{x\} \succ \{y\}$" to denote $\{y\}$ is majorized by $\{x\}$, meaning that the sum of the first $r$ elements in $\{x\}$ is no less than that in $\{y\}$ for any $r=1,\cdots$.
\end{definition}

Then we have the following two results:

\begin{lemma}
$\tilde{\lambda}_1 \geq \lambda_1$
\end{lemma}

{\em Proof:} Since $\tilde{{\bf D}}^H\tilde{{\bf D}}$ is a real-valued and symmetric matrix, we directly have:
\begin{align}
\tilde{\lambda}_1 &\geq \left(\frac{1}{\sqrt{JQ}} [1 \cdots 1]\right) \tilde{{\bf D}}^H\tilde{{\bf D}} \left(\frac{1}{\sqrt{JQ}} [1 \cdots 1]\right)^T\label{eqn:lemma5_lambda}\\
&= \frac{1}{JQ} \sum_m\sum_n (\tilde{{\bf D}}^H\tilde{{\bf D}})_{m,n} \\
& \geq \frac{1}{JQ} \sum_m\sum_n ({\bf D}^H{\bf D})_{m,n} \label{eqn:lemma5_lemma4}\\
& = J = \lambda_1\label{eqn:lemma5_J}
\end{align}
where (\ref{eqn:lemma5_lambda}) is obtained due to the fact that any unit-norm vector, rotated by $\tilde{{\bf D}}^H\tilde{{\bf D}}$ and then projected back to itself, cannot be greater than its largest eigenvalue (otherwise, the fact that $\tilde{\lambda}_1$ is the largest eigenvalue will be violated). Moreover, (\ref{eqn:lemma5_lemma4}) follows from Lemma 4, and (\ref{eqn:lemma5_J}) follows from Lemma 1.
\hfill\QED

\begin{lemma}
$\{\tilde{\lambda}\} \succ \{\lambda\}$
\end{lemma}

{\em Proof:} First, it can be easily verified that $\sum_{r=1}^R {\tilde{\lambda}}_r = \Tr (\tilde{{\bf D}}^H\tilde{{\bf D}}) = JQ$ and $\sum_{i=2}^N {\tilde{\lambda}}_i = JQ - \tilde{\lambda}_1$. Then we have:
\begin{align}
\{\tilde{\lambda}\} &\succ \left\{\tilde{\lambda}_1, \frac{JQ - \tilde{\lambda}_1}{R - 1}, \frac{JQ - \tilde{\lambda}_1}{R - 1}, \cdots, \frac{JQ - \tilde{\lambda}_1}{R - 1} \right\}\label{eqn:lemma6_1}\\
&\succ \left\{\lambda_1, \frac{JQ - \lambda_1}{R - 1}, \frac{JQ - \lambda_1}{R - 1}, \cdots, \frac{JQ - \lambda_1}{R - 1} \right\}\label{temp_637}\\
&\succ \{\lambda\}.\label{lemma6_last}
\end{align}
In the derivation above, (\ref{eqn:lemma6_1}) is obtained due to the fact that the elements in $\{\tilde{\lambda}\}$ from the second are sorted in a descending order, whereas those in the right-hand side are equal. That is, $\sum_i^R\tilde{\lambda}_i\geq JQ$ results in $\sum_i^r\tilde{\lambda}_i\geq JQr/R$, whereas the sum of the elements in the right-hand side of (\ref{eqn:lemma6_1}) is $JQ$ as well, which implies that the sum of its first $r$ elements is always equal to $JQr/R$. Moreover, (\ref{temp_637}) follows from Lemma 5, and (\ref{lemma6_last}) follows from Lemma 1.
\hfill\QED

Finally, based on the results above, we can directly derive the following:
\begin{align}
\Tr((\tilde{{\bf D}}^H\tilde{{\bf D}})^{-1}) &= \sum_r \tilde{\lambda}_r^{-1}\\
& \geq \sum_r \lambda_r^{-1}\label{eqn:lemma6_final}\\
& = \Tr(({\bf D}^H{\bf D})^{-1})
\end{align}
where (\ref{eqn:lemma6_final}) comes from Lemma 6 and the Schur convex property \cite{Marshall2011Inequalities}. This completes the entire proof of Theorem \ref{theorem_1}.
\hfill\QED

\subsection{The Proof of Proposition 2}\label{app-1}
When C1 is satisfied, the Gram matrix ${\bf \Gamma}$ of a codebook ${\bf P}(J, Q, K)$ is defined as:
\small
\begin{align}
{\bf \Gamma} = {\bf P}^H {\bf P}=\begin{bmatrix}
   {\bf I}_K & {\bf P}_1^H{\bf P}_2 & \dotsb & {\bf P}_1^H{\bf P}_J\\
   {\bf P}_2^H{\bf P}_1 & {\bf I}_K & \dotsb & {\bf P}_2^H{\bf P}_J\\
   \vdots & \vdots & \ddots & \vdots\\
   {\bf P}_J^H{\bf P}_1 & {\bf P}_J^H{\bf P}_2 & \dotsb & {\bf I}_K\\
   \end{bmatrix}_{JK \times JK}.
\end{align}\label{Gram_matrix}
\normalsize
Since $\Tr({\bf \Gamma})$ equals the sum of its eigenvalues, denoted by $\delta_q$, $q = 1, \dotsb, Q$, we can apply the Cauchy-Schwarz inequality to obtain
\begin{equation}
(\text{Tr} ({\bf \Gamma}))^2 = (JK)^2 =\Big(\sum_{q=1}^Q \delta_q\Big)^2 \leq Q\sum_{q=1}^Q \delta_q^2. \label{square_of_trace}
\end{equation}
Moreover, the square of the Frobenius norm of ${\bf \Gamma}$ satisfies
\begin{eqnarray}
\| {\bf \Gamma} \|^2_{\textrm{F}} &\!\!\!\!=\!\!\!\!& \sum_i\sum_j \|{\bf P}_i^H {\bf P}_j\|^2_{\textrm{F}} \\
&\!\!\!\!=\!\!\!\!& \sum_i\sum_{j\neq i} \|{\bf P}_i^H {\bf P}_j\|^2_{\textrm{F}} + JK = \sum_{q=1}^Q \delta_q^2. \label{F_norm}
\end{eqnarray}
Adding up (\ref{square_of_trace}) and (\ref{F_norm}) results in
\begin{align}
\sum_i\sum_{j\neq i} \|{\bf P}_i^H {\bf P}_j\|^2_{\textrm{F}} &\geq \frac{(JK)^2}{Q} - JK. \label{lower_bound2}
\end{align}
Owing to the condition C2, we have:
\begin{eqnarray}
\|{\bf P}_i^H {\bf P}_j\|^2_{\textrm{F}} = K^2c^2.\label{lower_bound1}
\end{eqnarray}
Substituting (\ref{lower_bound1}) into (\ref{lower_bound2}), we obtain:
\begin{align}
\sum_i^J\sum_{j=1,j\neq i}^J K^2c^2 &=(J^2-J)K^2c^2 \geq \frac{(JK)^2}{Q} - JK\\
&\Longrightarrow  c \geq \sqrt{\frac{JK - Q}{KQ(J - 1)}}.
\end{align}
Thus, we complete the proof.
\hfill\QED


\end{document}